\documentstyle[preprint,aps,psfig]{revtex}
\begin{document}

\draft
%{\tighten
\tightenlines
\preprint{\vbox{\hbox{U. of Iowa 97-2503; CERN 97-378}}}
\title{Accurate Checks of Universality for Dyson's Hierarchical Model}
\author{J. J. Godina\\
{\it Dep. de Fis. , CINVESTAV-IPN , Ap. Post. 14-740, Mexico, D.F. 07000\\
and\\
Dpt. of Physics and Astr., Univ. of Iowa, Iowa City, Iowa 52242, USA}}
\author{Y. Meurice\cite{byline} and M. B. Oktay\cite{byline}\\ 
{\it CERN, 1211 Geneva 23, Switzerland\\ 
and\\ 
Dpt. of Physics and Astr., Univ. of Iowa, Iowa City, Iowa 52242, USA}}
\maketitle

\begin{abstract}

Using recently developed methods, we perform high-accuracy
calculations of the susceptibility near $\beta _c$
for the $D=3$ version
of Dyson's hierarchical model. Using 
linear fits, we estimate the leading ($\gamma$) and subleading 
$(\Delta)$ exponents.
Independent estimates are obtained by
calculating the first two eigenvalues 
of the linearized  renormalization group transformation.
We found $\gamma = 1.29914073\pm10^{-8}$ and, 
$\Delta=0.4259469\pm 10^{-7}$ independently of the choice of
local integration measure (Ising or
Landau-Ginzburg). 
After a suitable rescaling, the approximate fixed points
for a large class of local measure
coincide accurately with 
a fixed point constructed by Koch and Wittwer. 
\end{abstract}

\narrowtext
\newpage

Scalar field theory has many important applications such as
superconductivity, low energy descriptions of quark-antiquark bound
states and possibly the mechanism responsible for the mass generation
of all the experimentally observed particles. However,
there exists no approximate treatment of this theory which could pretend
to compete in accuracy with  quantum electrodynamics at low energy,
where renormalized perturbation theory can be used to calculate 
the magnetic moment of the electron and the muon with
more than ten significant digits. Given the inherent difficulties associated 
with the experimental probing of very short distances, accurate calculations
compared with accurate ``low-energy'' 
experiments may become more of a standard procedure in 
the next century.

In order to obtain a completely satisfactory treatment 
of scalar field theory in various dimensions, 
one needs an approximation scheme such that: a) the
zeroth-order approximation preserves the main qualitative features 
of the model, b) the zeroth-order approximation allows
very accurate calculations, and c) the zeroth-order approximation 
can be improved systematically and in a way which
preserves its computational advantages.
We advocate here that hierarchical approximations such as 
the approximate recursion formula
derived by K. Wilson\cite{wilson} or the related recursion formula
which holds for Dyson's hierarchical model\cite{dyson} could be used as such a
zeroth-order approximation.
The fact that the approximate recursion formula 
satisfies the requirement a) is discussed at length in Ref.\cite{wilson}.
The fact that requirement b) is satisfied in the symmetric
phase is explained in Refs.\cite{bg,finite} and briefly reviewed below.
The use of the hierarchical approximation
solves some important problems encountered in practical
calculations in lattice field theory. First, it possible to perform 
all the integrals appearing in the calculation of the zero-momentum
Green's function in a much more efficient way than with the Monte-Carlo
procedure. Second, the computing time scales only like the log of the number
of sites, and one can eliminate finite-size effects completely. 
This can justify the effort of trying to solve 
part c) of the above program which is a very difficult problem.

Wilson's approximate recursion formula is closely related to the recursion
formula appearing in Dyson's hierarchical model\cite{dyson}. 
Both models have no wave function renormalization $(\eta=0)$.
It is possible to continuously interpolate between the two models and to show
that during this process, 
the critical exponent associated with the susceptibility 
$\gamma$ varies\cite{fam} by less than
5 percent with respect to the nearest neighbor value.
However, the numerical treatment of the two models is completely
identical.
In the following, we specialize the discussion
to the case of Dyson's model because this model
has been studied\cite{bg,sinai,epsi,wittwer,osc,high} 
in great detail in the past.
We want to make clear that this choice is not essential and anything done below
could have been done for Wilson's approximate formula.

In a typical lattice field calculation, we pick some values for 
the bare parameters entering in an action and we calculate the renormalized
quantities. In general, the physical masses cannot be too large
when expressed in cut-off units. Ideally, we should be able 
to cover a broad range of situations going from effective theories with
a low cut-off (e.g., $(m_{\rho}/m_{\pi}) \simeq 6$ for an effective theory
of pions) to a ``fundamental'' theory with a large cutoff and which
requires some fine-tuning procedure.
In the following calculations, 
the fixed bare parameters will appear in a local measure
of the Landau-Ginzburg (LG) form:
\begin{equation}
W_0(\phi)\propto exp^{-({1\over 2}m^2 \phi^2+ g\phi^{2p})} \ .
\label{eq:lg}
\end{equation}
The limit\cite{wilson}  of a large UV cut-off $\Lambda$, in units 
of the physical mass $m_R$, 
can be reached by tuning another parameter $\beta$, which is the 
inverse temperature
in Dyson's formulation of the model\cite{dyson}. More explicitly,
$\Lambda/m_R$ is proportional to $ (\beta_c-\beta)^{-{\gamma\over 2}}$.

By fine-tuning $\beta$, one can approach a fixed point of 
the renormalization group (RG) transformation and describe the RG flows
using the linear approximation.
As we will show later, for $\beta $ close 
enough to $\beta_c$ (i.e. , for $\Lambda$ large enough), 
one can approximate very well the 
magnetic susceptibility (zero-momentum two point function) with
a linearized expression taking into account only the first irrelevant 
direction:
\begin{equation}
\chi\simeq (\beta _c -\beta )^{-\gamma } (A_0 + A_1 (\beta _c -\beta)^{
\Delta } )\ ,
\label{eq:param}
\end{equation}
If we were sure that there is only one non-trivial fixed point 
(universality) and if we 
we could calculate accurately the exponents, then the 
complicated procedure described above can be reduced to
the determination of $A_0$
and $A_1$ in Eq.(\ref{eq:param}), a procedure that involves
no fine-tuning. 

In this Letter, we provide empirical evidence suggesting
that the RG transformation of Dyson's hierarchical model 
has only one non-trivial fixed point. We calculate the 
exponents $\gamma$ and $\Delta$ with two independent
methods (direct fit and linearization). 
The accuracy of our results is significantly better than 
the accuracy reached in the past \cite{bg,sinai,epsi,osc}.
All the 
approximate fixed points we have constructed below 
are very close (after 
rescalings explained below) 
to the fixed point calculated with an extraordinary
accuracy by Koch and Wittwer\cite{wittwer}. 
Our work demonstrates the enormous calculational advantage of 
using the hierarchical
approximation and addresses the question of understanding  
to what extent expansions about a known fixed point can be used 
as a substitute to the lengthy calculations in terms of bare parameters
described above.

For the sake of completeness, we briefly review the steps
which lead to the basic expression of the RG transformation of 
Eq.(\ref{eq:alg}). The
block-spin transformation of the hierarchical model is an
integral formula which transforms the {\it local} measure
$W(\phi)$ according to the rule: 
\begin{equation}
W_{n+1}(\phi)\propto  e^{{\beta \over 2} ({c\over 4})^{n+1} \phi ^2}
\int d\phi ' W_n({{(\phi -\phi ')}\over 2})
W_n({{(\phi +\phi ')}\over 2}) \ ,
\label{eq:bsp}
\end{equation}
where $c=2^{1-{2\over D}}$ in order to approximate $D$-dimensional nearest
neighbor models. 
For more details, the reader may consult Refs.\cite{finite,wittwer,osc}.
In the following we only consider the case $D=3$ in the symmetric phase.
We approach criticality for a fixed initial $W_0$ by fine-tuning $\beta $
as described in Ref. \cite{finite}.
When a critical value $\beta_c$ is reached approximately, a (discrete) scale
invariance is temporarily restored and it is convenient to reabsorb  
the scale factor $(c/4)$ in $\phi^2$. After this rescaling, 
we obtain the conventional
RG transformation of the local measure. In 
Fourier form it reads:
\begin{equation}
R_{n+1}(k)=C_{n+1}\exp(-{1\over 2}\beta 
{{\partial ^2} \over 
{\partial k ^2}})(R_{n}({k\sqrt{c}\over 2}))^2 \ . 
\label{eq:rec}\end{equation}
We fix the normalization constant $C_n$ in such way
that $R_n(0)=1$. We consider the finite dimensional
approximations of degree $l_{max}$:
\begin{equation}
R_n(k)=1+a_{n,1}k^2 +a_{n,2}k^4+..... +a_{n,l_{max}}k^{2l_{max}} \ .
\label{eq:rn} 
\end{equation}
The coefficients $a_{n,l}$ are proportional\cite{finite,high,osc} 
to the expectation value of the sum of all the 
fields (after $n$ iterations, there are $2^n$ of them) denoted $M_n$. 
In particular, the finite volume susceptibility $\chi_n (\beta)$,
defined as ${<(M_n)^{2}>_n /{2^{n}}}$ is simply $-2 a_{n,1} ({2\over c})^n $ .
When $\beta < \beta_c$, $\chi_n$ reaches a finite limit $\chi$ when $n$
goes to infinity.
The recursion formula for $a_{n,m}$ is purely algebraic:
\begin{equation}
a_{n+1,m}=
{{\sum_{l=m}^{l_{max}}(\sum_{p+q=l} a_{n,p}a_{n,q})
{{(2l)!}\over {(l-m)!(2m)!}}
({c\over 4})^l
(-{1\over 2}\beta)^{l-m}}\over
{\sum_{l=0}^{l_{max}}(\sum_{p+q=l} a_{n,p}a_{n,q}){{(2l)!}\over {l!}}
({c\over 4})^l
(-{1\over 2}\beta)^{l}}} \ .
\label{eq:alg}
\end{equation}
The initial condition for the 
Ising measure is
$R_0=cos(k)$. For the Landau-Ginsburg measure, the coefficients in the 
$k-$expansion need to be  evaluated numerically.

In a recent article\cite{finite},we have shown
that the errors on $\chi$ due to finite volume
and finite truncations fell 
exponentially fast with, respectively,
the number of iterations used and the dimension of 
the truncated space ($l_{max}$).
It is possible to make calculations where these errors 
play no practical role. The main limitation of the method comes from 
the round-off errors which are amplified when many iterations are spent
near the fixed point.
If the arithmetic operations are performed with a precision $\delta$, 
then\cite{finite} 
\begin{equation}
|{{\delta \chi }\over{\chi}}|\sim {{ \delta}\over{\beta _c - \beta}} \ .
\label{eq:numer}
\end{equation}

\begin{figure}
\centerline{\psfig{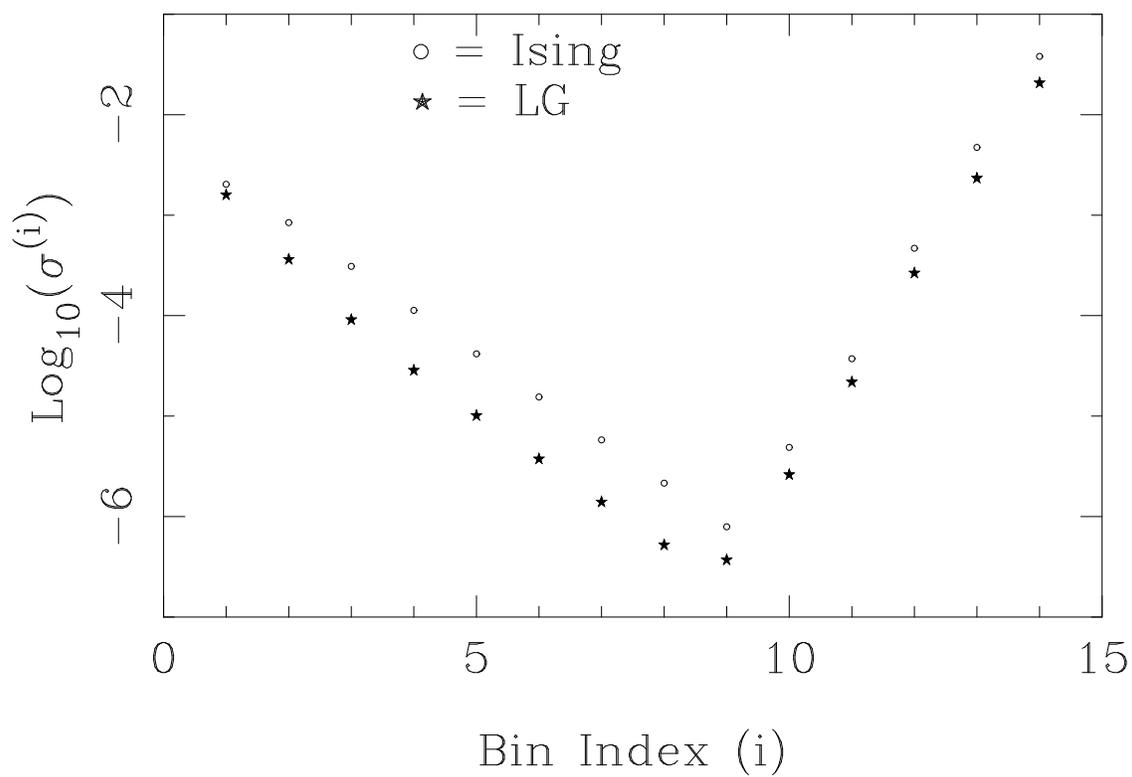}}
\vskip10pt
\caption{The deviations from the linear fits 
$\sigma ^{(i)}$ defined in the text as functions of the bins,
for the Ising model (circles) and the Landau-Ginzburg model (stars).}
\label{sig}
\end{figure}

We now proceed to determine the values of the four parameters appearing 
in Eq. (\ref{eq:param}) from direct calculations of $\chi$ at various 
temperatures. 
The calculations which follow have been performed 
for two particular choices of $W_0$, one corresponding to the 
Ising limit ($W_0(\phi)=\delta(\phi^2 -1)$) and the other 
to the choice $m^2=1,\ p=2$, and $g=0.1$ in Eq.(\ref{eq:lg}).
Unless specified differently, the calculations are performed 
using double-precision.
In the following, we use the 
notation $x$ for the quantity $-log_{10}(\beta_c -\beta)$.
If we display $log(\chi)$ versus $x$,
the deviations from the linear behavior are not visible to the eye
and need to be studied and understood ``locally'' in $\beta $.
In order to get a rough understanding of the corrections,
we have divided the computer data in 14 bins of 100 points.
The first bin contains data for values of $x= 1.00, 1.01,\dots 1.99$ 
and so on. In each bin (indexed $i$), 
we make a linear fit of $log_{10}(\chi)$ versus
$x$. 
In the $i$-th bin, we call the slope $\gamma ^{(i)}$, and 
$(\sigma ^{(i)})^2$ denotes the sum of the squares of the difference 
between the data and the linear fit divided by the number of points 
in a bin (100)
minus 2. 
The values of $\sigma ^{(i)}$ are displayed 
in Fig. \ref{sig}. This graph can be interpreted easily.
There are two known sources of deviations from the linear  behavior:
the subleading corrections to the scaling laws
(which decrease when $\beta $ gets close to $\beta_c$) and 
the round-off errors (which
increase when $\beta $ gets close to $\beta_c$ according to 
Eq.(\ref{eq:numer})). The approximate slopes in Fig. \ref{sig}
confirm this interpretation.
In bin 9, we minimize the combined deviations from linearity 
and we can
consider $\gamma^{(9)} $ as a first estimate of $\gamma$. 
Its numerical value is  1.29917 in the Ising case
and 1.29914 in the LG case.
With this simple-minded procedure, we have
already gained almost two significant digits compared to the existing
estimates\cite{bg,epsi,osc}
where the answer $\gamma=1.300$ was consistently obtained with
errors of order 1 in the last digit. 

We can improve this result by estimating the subleading
corrections. For this purpose, we have used the bins 6 and 7 
where the next subleading corrections are small (see discussion later) and
where the numerical errors are still not too large. We have divided these two
bins into 
10 sub-bins of 100 points each. We use two digit indices for these 
sub-bins. For instance sub-bin 6.3 is the third sub-bin of bin 6 and 
contains the values of $x$: $6.3, 6.301,\dots , 6.399$.
Using Eq.(\ref{eq:param}), the same 
kind of notations as above for $\gamma$ 
and noting that $ j+0.0495$ is the middle
of the sub-bin indexed by $j$, we obtain the approximate decay law:
\begin{equation}
\gamma^{(j)}\simeq \gamma -\Delta ({A_1\over A_0})10^{-\Delta(j+0.0495)} \ .
\end{equation}
The unknown coefficients can be extracted from linear fits of $log_{10}(
\gamma^{(j+0.1)}-\gamma^{(j)})$. We obtained
${A_1\over A_0}=-0.57$ and $\Delta=0.428$ for the Ising model
and ${A_1\over A_0}=0.14$ and $\Delta=0.427$ for the LG model specified above.
Repeating the first step (a linear fit in bin 9) 
but with $\chi$ divided by $(1+{A_1\over A_0}
(\beta_c-\beta)^{\Delta})$, we obtain $\gamma$=1.299141 with an agreement 
up to the sixth decimal place between the two models considered above.

Eq. (\ref{eq:numer}) is an unavoidable limitation if we use double precision
arithmetic. However,
using $Mathematica$ with a suitably set precision, $l_{max}=42$ for the 
Ising model and $l_{max}=50$ for the LG model (see ref.\cite{finite}
for the determination of these quantities), we were able to calculate $\chi$
in bins 10,11 and 12 with 11 significant digits. 
In the following, we call this data the ``high-precision data''.
Since this procedure is 
relatively lengthy, we have used only ten points per bins. We also determined 
$\beta_c$ with 24 significant digits so that in bin 12, 
the subtracted quantity
$\beta_c-\beta$ is also known with at least 11 significant digits.
In the Ising case, the result is easily reproducible and reads $\beta_c$=
1.17903017044626973251189. We have then used bin 12 
(where the subleading corrections are very small and our errors on them 
are less important) with $\chi$ divided by the subleading correction
as described above,
to estimate $\gamma $.
We then used this better value of $\gamma $ to obtain
the subleading corrections in bin 7 (where they are more sizable).
This procedure can be iterated.
This ``bootstrap'' of linear fits converges rapidly. We reach a nine
significant digit agreement between
the high-precision data and the fit obtained with the above procedure.
The small discrepancies can be analyzed in terms of first order
errors made in the estimate of the four parameters. 
This linear analysis provides small corrections ($<4 \times 10^{-9}$) 
to $\gamma$ and 
more sizable corrections $(<4\times 10^{-4})$ to $\Delta$.
The size of these corrections provide an order magnitude estimate
for the errors.
After these small corrections are taken into account, 
we obtain an agreement between the 
exponents of the two models for the following digits:
$\gamma=1.299140730$ and $\Delta=0.4260$. We conclude that 
$\gamma=1.29914073$ with an estimated error of less than $10^{-8}$.

We would like to comment about the corrections to Eq. (\ref{eq:param})
and how they could affect our estimates.
First, since the third eigenvalue of the linearized 
RG transformation $\lambda_3\simeq 0.48$, the next subleading exponent is
approximately 2. For $x>10$, these effects are negligible.
Second, a general argument\cite{wilson}, suggests that we should replace
the constant $A_0$ and $A_1$ in Eq.(\ref{eq:param}) 
by log-periodic function which can be expressed
as linear superposition of 
Fourier modes of the form 
$(\beta _c -\beta)^{i2\pi l \over ln(\lambda _1 )}$, with $l$ 
an integer.
Evidence for non-zero Fourier modes were found in Ref.\cite{osc} by
using an estimator of $\gamma -1$ called the extrapolated slope and 
denoted $\widehat{S}_m$. In this estimator, oscillating and constant 
contributions have roughly the same amplitude. However, using Eqs.(3.7) to 
(3.10) of Ref. \cite{osc}, one realizes that in $\widehat{S}_m$,
the oscillating amplitude is dramatically amplified by
a factor of the order $|\omega^3/\Gamma(\gamma+i\omega)|$
where $\omega=2\pi/ln(
\lambda_1)\simeq 18$. 
This implies that the Fourier coefficients of the non-zero
modes are suppressed by at least 14 orders of magnitude.
A direct search for these oscillations confirms this upper bound.
Third, Eq. (\ref{eq:param}) 
is obtained from a
linearization. Higher order
corrections give contributions proportional to 
$(\beta_c -\beta)^{2\Delta}$. An analysis of the difference between fit
and data in low bins indicates that these corrections
are the main source of errors in our analysis.

An alternative calculation of the exponents consists in linearizing
the RG transformation near a fixed point. An approximate fixed point
can be found by approaching  $\beta_c$ from below with our best resolution
and iterating until $a_{n+1,1}/a_{n,1}$ takes 
a value which is as close as possible
to 1. In the present formulation, the linearized RG transformation is given
by  the $l_{max}\times l_{max}$ matrix
\begin{equation}
M_{l,m}={{\partial a_{n+1,l}}\over {\partial a_{n,m}}}
\label{eq:linear}
\end{equation}
evaluated at the (approximated) fixed point. Using 
the high-precision $Mathematica$-based method described above,
we obtained this approximate fixed point for $n=101$ for the Ising model
and for $n=97$ for the LG model. Calculating the eigenvalues 
of Eq.(\ref{eq:linear}) for the two models used for the first estimates,
we obtain discrepancies of $2\times 10^{-8}$ for $\lambda_1$ and of
$4\times 10^{-8}$ for $\lambda_2$. The average values are
$\lambda_1=1.42717246$ and $\lambda_2= 0.85941163$. Changing $n$ by one or
improving the fixed point using Newton's method produce variations in these
eigenvalues which are smaller than $3\times 10^{-8}$.
Using the
relations $\gamma=2ln(2)/3ln(\lambda_1)$ and 
$\Delta=-ln(\lambda_2)/ln(\lambda_1)$,
we obtain $\gamma=1.29914078$ 
and $\Delta=0.4259469$ both with estimated errors of order $10^{-7}$.
The new estimate of $\gamma$ is compatible with the previous one but
is less accurate. On the other hand, the new estimate of
$\Delta$ is more accurate. The discrepancy with the previous estimate
is less than $10^{-4}$ which is compatible with our previous error estimate.

The two approximated fixed points obtained in the above calculation
depend on $\beta_c$. We denote them $R^{\star}(k,\beta_c)$.
However, it is possible to obtain what will turn out to be a universal 
function $U(k)$
by absorbing $\beta$ into $k$. More explicitly, we found that 
\begin{equation}
U(k)=R^{\star}(\sqrt{\beta_c}k,\beta_c)
\label{eq:ufonc}
\end{equation}
is in very good approximation independent of the model considered.
This function is related to a fixed point $f(s^2)$ constructed 
in Ref.\cite{wittwer}
by the relation
\begin{equation}
U(k)\propto f(({{c-4}\over{2c}})k^2)  \ .
\label{eq:trans}
\end{equation}
The Taylor coefficients of $f$ can be found in the file \verb+approx.t+
in\cite{wittwer}. Normalizing Eq.(\ref{eq:trans}) 
with $U(0)=1$, we obtain
\begin{equation}
U(k)=1. - 0.358711349882 k^2 + 0.053537288227 k^4 -\dots
\label{eq:ufp}
\end{equation} 
It is not known if there is only one non-trivial fixed point for 
Dyson's model.
Both the two approximate fixed points we have constructed above give 
a function $U(k)$ very 
close to Eq. (\ref{eq:ufp}). The closeness can be characterized by 
the $\rho$-norms introduced in 
\cite{wittwer}. For $\rho =2$
and $l\leq 42$ we found that the error 
$\delta u_l$ on the $l$-th coefficients of the 
approximate $U(k)$ with respect to the accurate 
expression obtained from Ref.\cite{wittwer} were bounded by 
$|\delta u_l|<{{3\times 10^{-8}}\over {l !2^l}}$.

In order to further explore the possibility of having different 
fixed points, we have considered more LG models. Using the parametrization
of Eq. (\ref{eq:lg}), we have considered the 12 cases obtained
by choosing among the following possibilities: $m^2=\pm 1 
$ (single or double-well potentials), $p=2,3$ or 4 (coupling constants
of positive, zero and negative dimensions when the cut-off is  restored)
and $g=10$ or 0.1 (moderately large and small couplings).
These searchs have been performed using regular double-precision calculations.
We have not aimed at great accuracy. For all these twelve models, we found
that using the same notations and conventions as a above 
$|\delta u_l|<{{5\times 10^{-5}}\over {l!2^l}}$.
In other words, the function $U(k)$ 
seems to be independent of the
general shape of the potential, the strength of the interactions
and whether or not the model is perturbatively renormalizable.

In conclusion, our best estimates of the critical exponents
$\gamma = 1.29914073\pm10^{-8}$ and, 
$\Delta=0.4259469\pm 10^{-7}$ have an accuracy significantly better 
than existing estimates\cite{bg,epsi,osc}.
Our results demonstrate the power of the methods developed 
in Ref. \cite{finite}. They provide an incentive to develop more efficient
perturbative calculations of the critical exponents and to attack
the problem of the improvement of the hierarchical approximation.
We found no indications for the existence of a non-trivial
fixed point different from the one obtainable from Ref. \cite{wittwer}.
Near criticality, or in field theoretical language for a large UV cut-off, 
the parametrization of Eq.(\ref{eq:param})
fits the data very well. 
The quantities $A_0$ and $A_1$ depend on the bare parameters 
in a complicated way. However, the fact that we can use confidently
the universal features suggests that it is possible to 
shortcut the use of bare parameters and consider directly 
$A_0$, $A_1$ as an input. 
More generally, we are in position to check if the following conjecture
is true: an expansion about the non-trivial fixed point can be used as 
a substitute for the calculations in terms of bare parameters.
If true, this 
would mean that the result of Ref.\cite{wittwer} effectively
``solves'' the hierarchical model even far away from criticality. 

This research was supported in part by the Department of Energy
under Contract No. FG02-91ER40664.
J.J. Godina is supported by
a fellowship from CONACYT. 
Y.M. thanks P. Wittwer and the CERN lattice group 
for useful conversations.

%} 
\end{document}